\documentclass[11pt]{article}
\usepackage[a4paper,margin=2.5cm]{geometry}
\usepackage[T1]{fontenc}
\usepackage{lmodern}
\usepackage{setspace}
\usepackage{lineno}
\usepackage{amsmath,amssymb,amsthm}
\usepackage{graphicx}
\usepackage{booktabs}
\usepackage{bm}
\usepackage[round]{natbib}
\usepackage[figurename=Fig.,labelfont=bf,labelsep=period]{caption}
\usepackage[colorlinks=true,citecolor=black,linkcolor=black,urlcolor=black]{hyperref}
\usepackage{xurl}

\newtheorem{theorem}{Theorem}
\newtheorem{proposition}{Proposition}
\newtheorem{corollary}{Corollary}
\theoremstyle{definition}
\newtheorem{assumption}{Assumption}
\newtheorem{remark}{Remark}

\newcommand{\Prob}{\mathbb{P}}
\newcommand{\E}{\mathbb{E}}
\newcommand{\R}{\mathbb{R}}
\newcommand{\cA}{\mathcal{A}}
\newcommand{\cX}{\mathcal{X}}
\newcommand{\cI}{\mathcal{I}}
\newcommand{\qd}{\hat q_\delta}
\newcommand{\xh}{\hat x}
\newcommand{\ch}{\hat c}
\newcommand{\fck}{f_{\mathrm{ck}}}
\newcommand{\fhat}{\hat f_{\mathrm{c}}}
\newcommand{\COe}{CO\textsubscript{2}e}
\newcommand{\unitC}{kg\,\COe/m\textsuperscript{3}}
\newcommand{\ind}{\mathbf{1}}

\title{Releasing Low-Carbon Concrete Mixes on Model Predictions: A Case Study of Miscalibration at the Selected Mix and of Release Protocols with a Guaranteed Failure Rate}
\author{Neelesh S. Upadhye\\[2pt]
\small Department of Mathematics, Indian Institute of Technology Madras, Chennai 600036, India\\
\small Email: \href{mailto:neelesh@smail.iitm.ac.in}{neelesh@smail.iitm.ac.in}\quad ORCID: 0000-0003-0537-4928}
\date{}

\begin{document}
\maketitle

\begin{abstract}
\noindent A ready-mixed concrete producer who adopts a new low-carbon mix must release it before its 28-day compressive strength, which decides compliance, can be measured. We present a case study of this quality decision on 103 laboratory mixes from a public data set, over 300 replications, using established statistical tools. A Gaussian process strength model of the kind used in constrained Bayesian optimization predicted strength with a root mean squared error of 0.9~MPa, yet at the cheapest mix it rated compliant it stated a compliance probability of 0.985 while that mix complied in only 0.727 of replications. Choosing the most favorable mix selects the mixes whose strength the model overstates. We then apply weighted conformal prediction, with the likelihood ratio fixed by the design of the calibration tests, as a release protocol whose probability of releasing a strength-failing mix is bounded in finite samples without assuming the model correct, and compare it with code-style strength margins, reuse of tested recipes and the model-probability rule on the same replications. The failure-rate bound held at every carbon budget, but the protocol released a mix in only 14.6\% of cases, calibration on laboratory strengths did not certify delivered batches, and in an expected-loss comparison code-style margins or reuse of tested recipes were the better policies. The case study separates two goals often run together in practice, a guaranteed failure rate for a released product and a useful release policy, and gives five recommendations for releasing a product on a model prediction.
\end{abstract}


\section{Introduction}
Most of the embodied carbon of concrete lies in its portland cement clinker \citep{anderson2020}, and the established way to reduce it is to replace part of the cement with slag or fly ash. These materials slow strength development, so a lower-clinker mix risks falling short of its specified 28-day strength unless the rest of the mix compensates. The usual protection, a margin of mean strength above the specification, is paid for in cement and therefore in carbon. The deciding response also arrives late. Slump is measured at the truck, so an unworkable batch is rejected on the spot, but the 28-day strength of a delivered batch is known only after the concrete is in the structure, and a trial batch of a new recipe costs the trial and a four-week wait. The decision to release a new low-carbon mix is therefore a decision under uncertainty, taken before its acceptance test exists, on whatever evidence the producer holds. Statistical and machine-learning models of strength, in use since \citet{yeh1998}, increasingly supply part of that evidence, and the question for the producer is what failure rate such a release actually carries.

We compare four ways of making that decision on the same mixes. The first reads a probability off a fitted model and releases the cheapest mix whose Gaussian process probability of compliance is at least $1-\delta$. This is the logic of constrained Bayesian optimization, which multiplies an acquisition function by the model's probability that each constraint holds \citep{schonlau1998,gardner2014,gelbart2014}, so that the number which steers the search is then read as the probability that the recommended mix complies. The second requires the predicted mean strength to exceed the specified strength by a fixed margin in the manner of the codes: 8.3~MPa under ACI 318-11 without a qualifying record of production results, and 4~MPa under EN 206 for initial production \citep{aci318,en206}. The third reuses a recipe that has already passed a strength test, and the fourth is a release protocol assembled from weighted conformal prediction, which carries a finite-sample bound on the probability of releasing a failing mix. Neither code claims to cover a prediction for a mix that has never been cast, so the first two rules are hypothetical. We study them because they are how a model prediction would naturally be turned into a release decision.

The difficulty is not prediction accuracy. On the 103 mixes studied here the strength model predicts with a root mean squared error of 0.9~MPa, against a standard deviation of 7.8~MPa across mixes, yet the probability it states at the mix it selects is badly overconfident. The reason is selection. A rule that releases the cheapest mix rated compliant favors mixes whose strength the model overstates, so the stated probability is most optimistic exactly where it is used. This is the optimizer's curse of decision analysis \citep{smith2006}. \citet{smith2006} show that a correctly specified Bayesian model removes it, and we show the same for the feasibility probability. For a fixed set of mixes, however, the model is only an approximation, and the curse returns. Two further mechanisms matter when the model also guides the search for candidate mixes, and we treat them in the supporting information: a search that places no trials where it judges compliance unlikely never revisits that judgment, and requirements that share physics, such as strength and workability, are dependent, while the usual product of marginal probabilities treats them as independent.

This article is a case study. Its methods are established: Gaussian process regression, the feasibility probability of constrained Bayesian optimization, and weighted conformal prediction under covariate shift \citep{tibshirani2019}. What it offers is their careful application to one real release decision, reported in full, with four findings. (i) A diagnosis (Section~3): the realized failure rate was about seven times the stated one over mixes the model rates compliant with probability at least 0.90, and eighteen times at the mix it selected. (ii) A release protocol (Section~4) under which the probability of releasing a strength-failing mix is at most $\delta$ in finite samples, whatever the model or the search that produced it. It is weighted conformal prediction with the likelihood ratio fixed by the design of the calibration campaign, written for testing distinct mixes from a finite candidate pool, together with the calibration effort it needs, which can be computed before any specimen is cast. (iii) A comparison (Section~5) of all rules on the same replications and delivered batches, ending in an expected-loss comparison of release policies with a named fallback after abstention. (iv) A population result: the protocol certifies the population its calibration tests came from, so calibration on laboratory strengths does not certify delivered batches.

The findings are mixed, and we report them as they occurred. The error control held at every carbon budget, but the protocol rarely released a mix at tight budgets and does nothing for workability. Given the same calibration information, it did not recommend lower-carbon mixes than simply reusing the cheapest tested recipe, and with a stated fallback no release policy using it had the smallest expected loss. The practical lesson is the distinction these results expose between a guaranteed failure rate and a useful release policy. The article is organized as follows. Section~2 describes the mixes, the specification and the sampling design. Section~3 examines the model's stated probability where the rule acts and explains the gap. Section~4 describes the release protocol, and Section~5 reports the results on the 103-mix population. Section~6 gives recommendations for practice, limitations and conclusions.

\section{Data, Specification, and Sampling Design}
We use the Concrete Slump Test data \citep{yeh2007a}: 103 laboratory mixes of concrete with fly ash, blast-furnace slag and superplasticizer, made in one laboratory, each recording the masses per cubic meter of cement, slag, fly ash, water, superplasticizer, coarse aggregate and fine aggregate, with the measured slump, flow and 28-day compressive strength. \citet{yeh2007b} describes the first 78 mixes and their workability: fly ash and slag together replace up to 68\% of the binder, and the water--binder ratio ranges from 0.25 to 0.70. The remaining 25 mixes and the strength record are documented only in the repository, and neither source states the test specimen. We take a specified strength $\fck=35$~MPa, the cylinder strength of EN 206 class C35/45 and close to the common specification of 5,000~psi, which lies at the median of the observed strengths (35.5~MPa) and is met by 53 of the 103 mixes. We take a slump window $[S_{\mathrm{lo}},S_{\mathrm{hi}}]=[15,25]$~cm, roughly the central half of the observed slumps (median 21.5~cm), met by 62; 23 mixes meet both. The two requirements differ in when they can be checked. Strength is latent at release, and we write its requirement as $c_{\mathrm{lat}}(x)=\fck-f_{\mathrm{c}}(x)\le 0$. Slump is observed at the truck, with requirement $c_{\mathrm{obs}}(x)=\max\{S_{\mathrm{lo}}-S(x),\,S(x)-S_{\mathrm{hi}}\}\le 0$. Here $f_{\mathrm{c}}$ is the strength and $S$ the slump. Only the latent requirement is certified, so the protocol of Section~4 is applied with $c=c_{\mathrm{lat}}$ and $\ch=\fck-\fhat$ for the frozen strength model $\fhat$, and workability is left to the measurement at the truck. The objective is embodied carbon from factory-gate coefficients \citep{mpa2025,hammond2011,efca2015}. The cheapest compliant mix is the same under four inventory conventions, but carbon differences between mixes change by up to a factor of three across them (Section~S19 of the supporting information), so we report differences under the primary set and read them with that range in mind.

Each of 300 replications mimics a producer with a history. A uniform 40-mix development set DEV, standing for the producer's past trial batches, is drawn. Gaussian process models of strength and slump are fitted on it and frozen, and the remaining 63 mixes form the pool of candidates. Every procedure sees only what it would see in practice: the outcome of a pool mix is withheld until a procedure tests it. The mix offered for release, the report, is drawn uniformly from the region $\cA(\tau)$ of pool mixes with carbon at most $\tau$ and a model probability of at least 0.45 of meeting the slump window, a prespecified screening threshold set slightly below one half so that mixes the slump model is unsure about are not excluded before slump is measured at the truck. The region is fixed from the frozen models before any testing. The calibration campaign then tests $n_{\mathrm{cal}}=40$ of the remaining 62 pool mixes, drawn without replacement, which is the sampling law of Corollary~1 with $N=63$. This is a large campaign, and Section~5.3 charges for it. The nine carbon budgets $\tau$ are prespecified: the 5th, 10th, 20th, 30th, 40th, 50th, 65th, 80th and 100th percentiles of carbon over the 103 mixes. Pooled results are for the procedure that draws one budget at random (Remark~2).

The data record one strength per mix, so a delivered batch is simulated as $f_{\mathrm{c}}^{\mathrm{batch}}=f_{\mathrm{c}}^{\mathrm{lab}}+\varepsilon$ with $\varepsilon\sim N(0,s^2)$. We use $s=0$, the laboratory case, in which each rule is scored on the recorded strength, and $s=2$ and 4~MPa. Each mix and replication has one draw, shared by every rule. The Gaussian form belongs to the simulation, not to the guarantee. Two external records suggest that $s=2$~MPa is realistic and $s=4$~MPa a stress case. In a road-and-bridge contractor's mix-proportion ledger \citep{lu2024}, the within-mix standard deviation of 28-day strength is 1.8~MPa at a mean of 41~MPa, a coefficient of variation of 4.4\%. Across laboratory batches of one high-performance recipe \citep{rezazadeh2026}, it is 5.0\%. At 35~MPa these correspond to about 1.5 and 1.8~MPa (Section~S30 of the supporting information).

\section{The Model's Probability Where the Rule Acts}
\subsection{The Probability at the Selected Mix}
The strength model is a Gaussian process with a Mat\'ern 5/2 covariance and a separate length-scale for each ingredient, its hyperparameters fitted by maximum marginal likelihood on the development set. As a predictor it is accurate. Averaged over replications, its root mean squared error on the pool mixes is 0.90~MPa, against a standard deviation of 7.8~MPa across mixes, because the recorded strengths are close to a smooth function of the mix (Section~S18 of the supporting information). This is a low-noise regime, to which we return in Section~6.2. Table~\ref{tab:calib} compares the stated probability of compliance with realized compliance in three groups of pool mixes: (a) all mixes the model rates compliant with probability at least 0.90; (b) the 40\% of those whose predicted margin over $\fck$ is smallest; and (c) the single mix the model-probability rule selects, the cheapest member of (a).

\begin{table}[t]
\caption{Stated probability of compliance against realized compliance of the delivered batch}
\label{tab:calib}
\centering\footnotesize\setlength{\tabcolsep}{4pt}
\begin{tabular}{@{}lrrrrrr@{}}
\toprule
Group & $n$ & Stated & Realized & (SE) & Overstatement & Failure ratio\\
\midrule
\multicolumn{7}{@{}l}{Laboratory case, $s=0$}\\
(a) All mixes rated $\ge 0.90$ & 4,872 & 0.996 & 0.970 & 0.002 & $+0.026$ & 7.3\\
(b) Rated mixes nearest the boundary & 2,007 & 0.990 & 0.929 & 0.005 & $+0.061$ & 7.3\\
(c) The mix the rule selects & 300 & 0.985 & 0.727 & 0.026 & $+0.258$ & 18.4\\
\multicolumn{7}{@{}l}{Production variability $s=2$~MPa}\\
(a) & 4,872 & 0.996 & 0.935 & 0.003 & $+0.061$ & 16.1\\
(b) & 2,007 & 0.990 & 0.847 & 0.008 & $+0.144$ & 15.7\\
(c) & 300 & 0.985 & 0.670 & 0.027 & $+0.315$ & 22.2\\
\multicolumn{7}{@{}l}{Production variability $s=4$~MPa}\\
(a) & 4,872 & 0.996 & 0.865 & 0.005 & $+0.131$ & 33.5\\
(b) & 2,007 & 0.990 & 0.716 & 0.010 & $+0.274$ & 29.0\\
(c) & 300 & 0.985 & 0.597 & 0.028 & $+0.388$ & 27.2\\
\bottomrule
\end{tabular}

\smallskip
\parbox{\linewidth}{\footnotesize Note: 300 replications, $\delta=0.10$. Groups: (a) all pool mixes rated compliant with probability at least 0.90; (b) the 40\% of these with the smallest predicted margin over $\fck$; (c) the mix the model-probability rule selects. Rows (a) and (b) pool (replication, mix) pairs, with standard errors clustered by replication. The failure ratio is $(1-\text{realized})/(1-\text{stated})$.}
\end{table}

Over group (a) a stated 0.996 meets a realized 0.970. That is close in absolute terms, but the realized failure rate is 7.3 times the stated one. At the selected mix a stated 0.985 corresponds to a realized 0.727, a failure ratio of 18: roughly one released mix in four would fail. The gaps widen with production variability, and at $s=4$~MPa the selected mix complies in only 0.597 of replications. This is the optimizer's curse at work, and the model's own uncertainty does not explain it. Compare the selected mix with non-selected mixes that the frozen model describes almost identically, in predicted margin and predictive standard deviation. The selected mix still overstates its compliance by 0.115 to 0.135 more, against a raw difference of 0.248, with intervals clear of zero (Section~S21 of the supporting information). No such comparison can separate selection from ingredient composition, but for this model and selection rule the stated probability was not an error rate where the rule acts.

\subsection{Why the Stated Probability Is Not an Error Rate}
The gap has a general explanation, which we state for constrained Bayesian optimization and prove in the supporting information. Write $f,g_1,\dots,g_m:\cX\to\R$ for the objective and the constraints on a compact design space $\cX$, and
\begin{equation}
F=\{x\in\cX: g(x)\le 0\},\qquad x^\star\in\arg\min_{x\in F} f(x).
\end{equation}
With independent Gaussian process priors on the $g_j$, the usual criterion maximizes
\begin{equation}
\tilde\alpha_t(x)=\alpha_t(x)\prod_{j=1}^m \Prob\bigl(g_j(x)\le 0\mid D_t\bigr)
\end{equation}
after $t$ evaluations $D_t$, where $\alpha_t$ is an acquisition function such as expected improvement. The product, or the joint probability $\pi_t(x)=\Prob(g(x)\le 0\mid D_t)$, is then read as the probability that a design complies. In the concrete study $m=1$ for the certified requirement and $f$ is embodied carbon. The model-probability rule releases the cheapest pool mix with $\pi\ge 1-\delta$.

\emph{Calibration under a correct model.} If the constraints are drawn from the prior the model uses, then for every fixed $x$ the probability $\pi_t(x)$ is a bounded martingale in $t$, whatever adaptive rule chooses the evaluations (Proposition~S3), and at a design chosen from the data $\pi_t$ is still the conditional probability that it complies given the data. A correct model is therefore calibrated where the rule acts. This is the Bayesian remedy for the optimizer's curse that \citet{smith2006} describe.

\emph{Where it fails.} In our study the 103 strengths are fixed, and only the development set varies between replications. Calibration is therefore judged over the randomness of that design, which the prior does not describe, and the Gaussian process is only a working approximation to a fixed response. Its errors then need not be neutral where the rule acts: releasing the cheapest mix rated compliant favors mixes whose strength the model overstates. The same overconfidence on evaluated designs appeared in a tension-spring benchmark and in a groundwater remediation simulator, where better-fitting surrogates improved the objective but worsened calibration (Sections~S3, S4 and S7 of the supporting information).

\emph{Censoring and dependent requirements.} When the model also guides the search, a region it judges unlikely to comply receives no evaluations, so under a misspecified model the exclusion can be permanent (Proposition~S4). When several requirements must hold together, the product in Eq.~(2) understates the joint probability under positive posterior dependence (Proposition~S1), although correcting for dependence improved regret in only one of twelve prespecified settings (Section~S4.2). Neither issue arises in the concrete study: the candidate pool is finite and fully enumerated, and slump is measured at the truck, so only strength needs a statement made in advance. The remedy we pursue is to separate the two roles of the probability: let the model choose what to release, and attach to the released mix a statement checked against data that the choice did not use.

\section{A Release Protocol with a Finite-Sample Error Guarantee}
\begin{figure}[t]
\centering
\includegraphics[width=\linewidth]{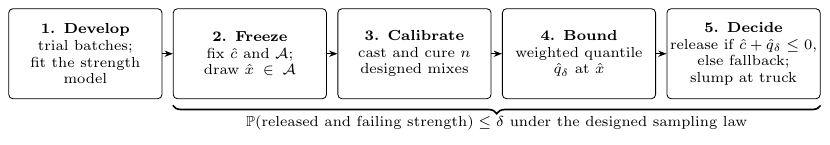}
\caption{The release workflow. Step 1 may use any model or search, including constrained Bayesian optimization; Steps 2--5 attach to the released mix a strength statement whose error rate is controlled in finite samples, whatever happened in Step 1.}
\label{fig:workflow}
\end{figure}

We now give a protocol that either releases a mix with a statement whose error rate is controlled in finite samples, without assuming that the model is correct, or abstains. Let $c$ be a scalar requirement, here $c=c_{\mathrm{lat}}$, with $c(x)\le 0$ meaning compliant, and let $\ch$ be a predictor of it produced by any earlier analysis or search. Two obstacles stand between the model and an error rate. The first is adaptivity: residuals of a predictor on its own training data are not exchangeable with those at a fresh mix. The second is selection: the released mix is chosen where the model is most favorable. Both are removed by designing the experiment for release. Fig.~\ref{fig:workflow} shows the resulting workflow.

\subsection{The Protocol}
Write $\cI_0$ for the $\sigma$-field generated by everything fixed before calibration: the development data, the fitted predictor, the candidate population and every design choice made in advance. Fix $\delta\in(0,1)$.

\begin{assumption}[Certification protocol]
(P1) \emph{Freeze.} The selection phase, by any procedure whatsoever, produces a predictor $\ch$ measurable with respect to $\cI_0$, which then does not change.
(P2) \emph{Calibrate.} Draw $X_1,\dots,X_n$ conditionally i.i.d.\ from an $\cI_0$-measurable $Q_{\mathrm{cal}}$, evaluate the constraint only, and form the residuals $E_i=c(X_i)-\ch(X_i)$.
(P3) \emph{Randomize the report.} Draw $\xh$ from an $\cI_0$-measurable report distribution $Q_{\mathrm{rep}}\ll Q_{\mathrm{cal}}$, conditionally independently of the calibration sample.
\end{assumption}

Because the experimenter chooses both $Q_{\mathrm{cal}}$ and $Q_{\mathrm{rep}}$, the likelihood ratio
\begin{equation}
w(x)=\frac{dQ_{\mathrm{rep}}}{dQ_{\mathrm{cal}}}(x)
\end{equation}
is known, and weighted conformal prediction under covariate shift \citep{tibshirani2019} is valid in finite samples with no estimated density ratio. The price is a report randomized over a region rather than a deterministic point (Section~S17 of the supporting information gives admissible constructions).

In the application the protocol takes a simple form. The models and the region $\cA(\tau)$ are fixed, the report $\xh$ is drawn uniformly from $\cA(\tau)$, and $n=40$ other pool mixes are cast and tested. For each tested mix inside $\cA(\tau)$ we record how far the model overpredicted its strength, $E_i=\fhat(X_i)-f_{\mathrm{c}}(X_i)$. With $n_\cA$ such mixes and $k=\lceil (n_\cA+1)(1-\delta)\rceil$, the bound $\qd$ is the $k$th smallest of these overpredictions, and it is infinite if $k>n_\cA$. The mix is released if its predicted strength, reduced by the bound, still meets the specification: $\fhat(\xh)-\qd\ge\fck$. Otherwise the protocol abstains. At $\delta=0.10$ this needs at least nine tested mixes in the region, and with exactly nine it subtracts the largest overprediction among them. Corollary~1 gives the guarantee for this form.

\subsection{The Certificate}
Define for a candidate report $x$ the masses $p_i(x)=w(X_i)/\{\sum_{k=1}^n w(X_k)+w(x)\}$ and the weighted conformal quantile
\begin{equation}
\qd(x)=\inf\Bigl\{r\in\R:\ \textstyle\sum_{i=1}^n p_i(x)\,\ind\{E_i\le r\}\ \ge\ 1-\delta\Bigr\},
\end{equation}
with $\qd(x)=+\infty$ when the set is empty.

\begin{theorem}[Finite-sample compliance certificate]
Let Assumption~1 hold and let $c$ be evaluated without noise. Then
\begin{equation}
\Prob\bigl(c(\xh)\le \ch(\xh)+\qd(\xh)\bigr)\ \ge\ 1-\delta,
\end{equation}
over the calibration and report draws jointly, conditionally on $\cI_0$, and hence also unconditionally. If the procedure issues the report only when $\ch(\xh)+\qd(\xh)\le 0$, it issues a noncompliant report with probability at most $\delta$, and
\begin{equation}
\Prob\bigl(c(\xh)\le 0\mid \text{a report is issued}\bigr)\ \ge\ 1-\delta\big/\Prob(\text{a report is issued})
\end{equation}
whenever $\Prob(\text{a report is issued})>0$. No assumption is made on $\ch$ or on the procedure that produced it.
\end{theorem}

Theorem~1 is the finite-sample guarantee of weighted conformal prediction \citep{tibshirani2019} written for this design, and Corollary~1 is its form for sampling without replacement from a finite pool. Proofs are in Section~S5 of the supporting information, with machine-checked counterparts in the replication package.

\begin{remark}[Procedure-level error control]
The theorem controls a property of the procedure across repetitions, the probability of issuing a noncompliant report, not the reliability of a particular issued report. The conditional bound (6) is vacuous whenever the issuance probability is at most $\delta$: at $\delta=0.10$ and the pooled issuance rate of 0.146 in Section~5.1 it guarantees only $1-0.10/0.146=0.315$. In practice: over many release decisions made in this way, at most a fraction $\delta$ release a mix that fails, but no particular released mix carries that guarantee.
\end{remark}

\begin{remark}[The procedure must be fixed in advance]
The guarantee attaches to a procedure specified in advance, including its report region. Fixing a carbon budget before calibration satisfies this, and so does drawing the budget at random from a prespecified list, whose error rate is the average of the per-budget rates; the pooled rows of Section~5 are that mixture. Adaptive use of the outcome, such as loosening the budget until a report issues or redrawing after an abstention, is a different procedure whose error rate is not bounded here.
\end{remark}

\subsection{Sampling without Replacement from a Finite Population}
On a finite candidate set, (P2) means sampling with replacement. Where each mix has one fixed strength, testing a mix twice gives the same residual and no new information, so the experimenter instead tests $n$ distinct candidates, none of them the report. The corresponding specialization uses the same rank argument; we state it because the application requires exactly this sampling law.

\begin{corollary}[Finite-population certification without replacement]
Let $G$ be a finite population of $N$ units and $\cA\subseteq G$ a nonempty region, both $\cI_0$-measurable. Draw the report $\xh\sim\mathrm{Unif}(\cA)$ and the calibration set as a uniform $n$-subset of $G\setminus\{\xh\}$, without replacement. Set $w_i=\ind\{X_i\in\cA\}$ and let $\qd$ be defined by Eq.~(4) with report weight 1. Then Eqs.~(5) and (6) hold for every $N$, $n\le N-1$ and $|\cA|$, with no continuity requirement on the residuals.
\end{corollary}

The calibration set must exclude the report, and an empty region is treated as an abstention, not as a certified report. A producer may object to releasing a mix drawn at random rather than the cheapest one. The randomization is what makes the error rate hold whatever the model, and every mix in $\cA(\tau)$ meets the carbon budget, so the cost is that the released mix is a random one within budget rather than the cheapest; Section~5 measures it. Corollary~S3 extends both results to a measured outcome, provided calibration and release use a common measurement law.

\subsection{The Cost of Certification}
Choosing the region from the frozen model is not the selection the certificate guards against, provided no calibration outcome is used (Corollary~S4). What is not free is precision: with weights constant on $\cA$, the quantile depends only on the calibration points inside it.

\begin{proposition}[Effective calibration size and vacuity threshold]
Let $n_\cA=\#\{i\le n: X_i\in\cA\}$. Then
\begin{equation}
\qd<\infty\iff n_\cA\ge\frac{1-\delta}{\delta},
\end{equation}
so at $\delta=0.10$ at least nine calibration specimens must lie inside the report region.
\end{proposition}

Shrinking the region therefore makes the certificate vacuous, never invalid. Under Corollary~1 the in-region count is hypergeometric, so
\begin{equation}
\E[n_\cA]=\frac{n(|\cA|-1)}{N-1},\qquad \Prob(\qd<\infty)=\Prob\bigl(n_\cA\ge(1-\delta)/\delta\bigr),
\end{equation}
both computable before any specimen is cast. In the study, with $n=40$ tests drawn from the 62 pool mixes other than the report, the expected in-region count reaches nine only when the region holds about 15 of the 63 pool mixes. Table~\ref{tab:cert} shows the consequence: at the five tightest budgets the median region holds at most 14 mixes, and the bound is finite in at most about half of the replications.

\subsection{Relation to Existing Conformal Methods}
The coverage argument is that of weighted conformal prediction \citep{tibshirani2019}; what the protocol adds is a design under which its hypotheses hold after an arbitrary search. The calibration tests are new and are made after the model is frozen, and the likelihood ratio is known by design rather than estimated or approximated as in \citet{barber2023}. Under feedback covariate shift \citep{fannjiang2022} the guarantee averages over a superpopulation, whereas here it conditions on everything the search produced; selection-conditional methods \citep{jin2025} need exchangeability between selected and calibration units, which this design gives up. The finite-population form follows design-based conformal inference \citep{wieczorek2023} in resting on the sampling design rather than on a superpopulation model.

\section{Results on the Concrete Population}
\subsection{Release Rules and the Strength-Compliance Certificate}
Because only strength is certified, the protocol issues a strength-compliance certificate, not a certificate of a feasible mix. If it releases a mix, the slump of the delivered batch is still measured at the truck, and the batch is accepted or rejected on that measurement.

Table~\ref{tab:loss} in Section~5.3 gives the failure rate and carbon of each rule at each level of production variability. The two code-style margins, applied to the predicted mean strength as a hypothetical use of the codes \citep{aci318,en206}, protect against production variability rather than prediction error. They never fail in the laboratory case. At $s=4$~MPa the EN-style rule complies in 0.910 of replications and the ACI-style rule in 0.987, but at a carbon cost: the mixes they release average 219.6 and 276.4~\unitC. (These averages are over the mixes released; Table~\ref{tab:loss} averages per decision, counting a decision with no release as zero carbon.) Reusing the cheapest recipe that passed in the development set complies in every replication in the laboratory case, at 169.1~\unitC. At $s=2$ and 4~MPa it complies in only 0.783 and 0.683 of replications: a recipe that passed once is not protected against batch-to-batch variation. The model-probability rule is the least reliable at every level, complying in 0.727, 0.670 and 0.597 of replications while stating 0.985.

\begin{table}[t]
\caption{The certificate in the laboratory case $s=0$, by carbon budget $\tau$ (\unitC)}
\label{tab:cert}
\centering\scriptsize\setlength{\tabcolsep}{1.7pt}
\begin{tabular}{@{}rrrrrrrrrrrrr@{}}
\toprule
 & & & & \multicolumn{5}{c}{Procedure} & \multicolumn{4}{c}{Among issued}\\
\cmidrule(lr){5-9}\cmidrule(l){10-13}
$\tau$ & med $|\cA|$ & med $n_\cA$ & Finite & \# iss. & Iss.\ rate & Coverage (SE) & \COe & Iss.\ \& fail & Compliant & 95\% interval & Workable & Usable\\
\midrule
147.3 & 3.0 & 1.0 & 0.000 & 0 & 0.000 & 1.000 (all)$^{a}$ & --- & 0.000 & --- & --- & --- & ---\\
160.1 & 5.0 & 2.0 & 0.000 & 0 & 0.000 & 1.000 (all)$^{a}$ & --- & 0.000 & --- & --- & --- & ---\\
166.2 & 8.0 & 5.0 & 0.080 & 0 & 0.000 & 0.993 (0.005) & --- & 0.000 & --- & --- & --- & ---\\
176.6 & 11.5 & 7.0 & 0.257 & 14 & 0.047 & 0.977 (0.009) & 168.2 & 0.003 & 13/14 & [0.661, 0.998] & 7/14 & 6/14\\
193.7 & 14.0 & 9.0 & 0.527 & 18 & 0.060 & 0.967 (0.010) & 165.6 & 0.013 & 14/18 & [0.524, 0.936] & 13/18 & 10/18\\
249.4 & 16.0 & 10.0 & 0.593 & 38 & 0.127 & 0.953 (0.012) & 184.1 & 0.003 & 37/38 & [0.862, 0.999] & 9/38 & 8/38\\
282.5 & 22.0 & 13.0 & 0.797 & 74 & 0.247 & 0.940 (0.014) & 230.1 & 0.010 & 71/74 & [0.886, 0.992] & 35/74 & 33/74\\
295.6 & 28.0 & 17.0 & 0.957 & 121 & 0.403 & 0.940 (0.014) & 256.0 & 0.000 & 121/121 & [0.970, 1.000] & 67/121 & 67/121\\
350.1 & 37.5 & 23.0 & 0.980 & 129 & 0.430 & 0.923 (0.015) & 271.4 & 0.020 & 123/129 & [0.902, 0.983] & 72/129 & 66/129\\
\multicolumn{8}{@{}l}{Uniform mixture over the nine budgets} & 0.0056 & 379/394 & [0.941, 0.980]$^{b}$ & & \\
\bottomrule
\end{tabular}

\smallskip
\parbox{\linewidth}{\footnotesize Note: Medians are over 300 replications; ``Finite'' is the fraction with a finite bound. ``Iss.\ \& fail,'' the probability of releasing a mix that fails strength, is the rate bounded by $\delta=0.10$ (Corollary~1); the columns among issued reports are not guaranteed. Coverage is Eq.~(5), which counts a vacuous bound or an empty region as covering; over the 2,677 cells with a nonempty region, pooled coverage is 0.966. $^{a}$All 300 covered (exact one-sided 95\% lower bound 0.990); at $\tau=147.3$ coverage is one by construction. Per-budget intervals are Clopper--Pearson; $^{b}$the pooled interval is a bootstrap over replications.}
\end{table}

\begin{figure}[t]
\centering
\includegraphics[width=\linewidth]{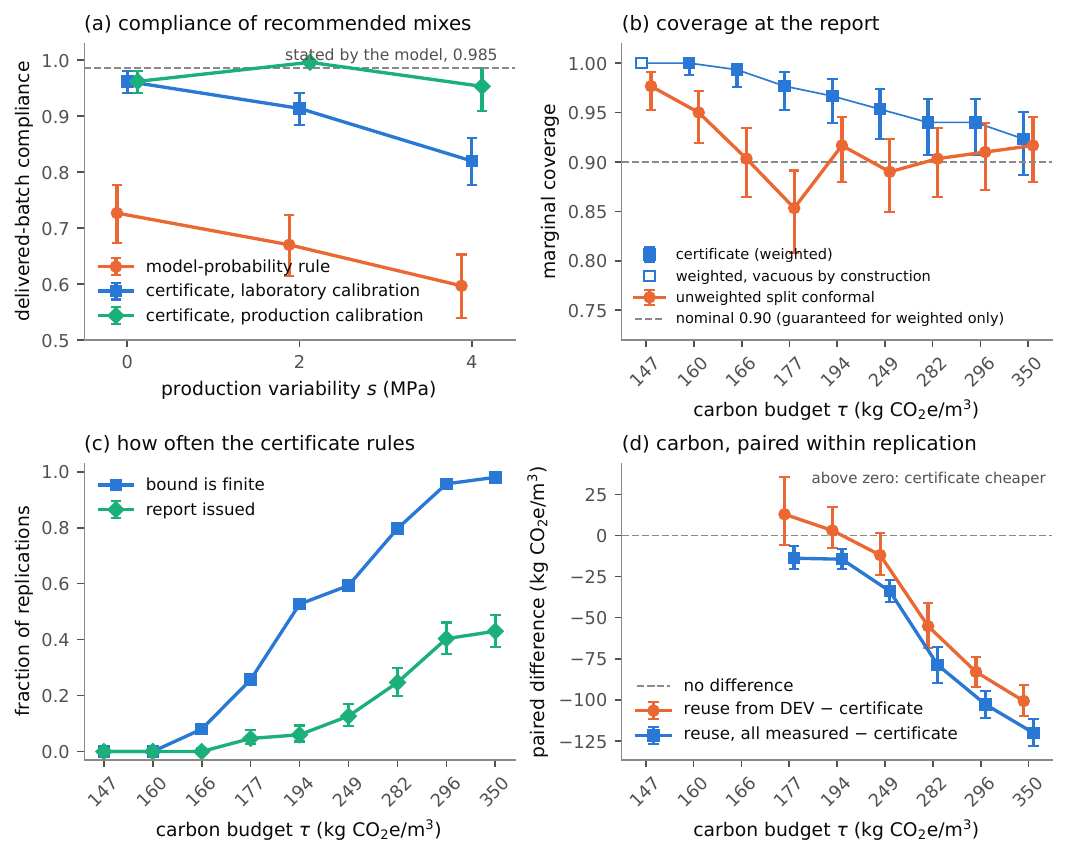}
\caption{(a) Delivered-batch compliance against production variability, with 95\% intervals; the dashed line is the probability the model states at its selected mix. Certificate points pool the budgets and are descriptive. (b) Marginal coverage with Clopper--Pearson intervals, denominator 300; the nominal 0.90 is guaranteed only for the weighted rule (open marker: one by construction). (c) Fraction of replications with a finite bound and with an issued report. (d) Carbon of each recipe-reuse rule minus the certificate, paired within issuing replications, with cluster-bootstrap intervals. Panels (b)--(d) are the laboratory case.}
\label{fig:results}
\end{figure}

Table~\ref{tab:cert} sweeps the carbon budget for the strength-compliance certificate. The guarantee holds. The bound held at the drawn mix in 0.923 (0.015) to 1.000 of replications against the nominal 0.90, and the rate of releasing a failing mix was at most 0.020 at every budget and 0.006 for the mixture. It rarely releases at tight budgets. Nothing was released at the three tightest budgets: at the two tightest, the region is empty or holds fewer than the nine tested mixes that Proposition~1 requires, and at $\tau=166.2$ the bound is finite in 8\% of replications but never clears the specification. Above that the protocol releases in 4.7\% to 43\% of replications, and in 14.6\% over the mixture, so in about six decisions out of seven the producer must fall back on another rule. Released mixes complied, but many were unusable. Of 394 released mixes, 379 met the strength specification, 0.962 [0.941, 0.980], against 0.727 for the model-probability rule. That is a measured frequency, not a guarantee: at the pooled release rate the conditional bound (6) guarantees only 0.315 (Remark~1). And only 48.2\% would also have passed the workability check at the truck. Fig.~\ref{fig:results} summarizes the study.

\subsection{Information, Population, and Calibration Design}
Three comparisons qualify the certificate's reliability; Tables~S15, S16 and S17 of the supporting information give the details. These comparisons condition on release; Section~5.3 compares policies with a fallback after abstention.

What does the certificate add to the information its campaign collects? Two of its ingredients matter. Without the weights, split conformal prediction on the same tests covers in only 0.853 of replications at $\tau=176.6$, below the nominal 0.90, and its released mixes comply in 0.724 to 0.828 of replications at the budgets from 166.2 to 193.7. Without the option to abstain, the same randomly drawn mix released without a certificate complies in only 0.160 to 0.483. But the campaign also tests 40 recipes. Reusing the cheapest recipe that passed among all tested mixes, in the development set and the campaign, complies in every replication in the laboratory case. Paired within the replications in which the certificate releases, it is cheaper by 13.8 [6.1, 20.4]~\unitC\ at $\tau=176.6$, and by more at looser budgets [Fig.~\ref{fig:results}(d)].

Which population does it certify? The one its tests come from. Calibrated on laboratory strengths, the protocol certifies laboratory strength. At $s=4$~MPa its bounded error rate stays at 0.006 and its coverage at or above nominal, while its released mixes comply on the delivered batch in only 0.820 of cases. Calibrating on production batches restores the delivered-batch guarantee, but the release rate falls from 14.6\% to 9.4\% and 5.5\% at $s=2$ and 4~MPa. The certificate's own coverage diagnostics cannot reveal the mismatch, because they are computed on the calibration population.

Can the campaign be designed better? Partly. Drawing the tested mixes from the release region itself, rather than from the whole pool, raises the frequency of a finite bound at the four middle budgets from 8--59\% to 37--87\%. It raises the release rate from 14.6\% to 18.6\%, with the same guarantee, and its error rate was at most 0.027 at every budget. With production calibration, the in-region certificate's released mixes comply in 0.972 and 0.948 of deliveries at $s=2$ and 4~MPa. By contrast, reusing the cheapest recipe whose single test batch passed complies on a fresh delivery in only 0.706 and 0.549.

\subsection{Strength-and-Carbon Loss of Release Policies}
A certificate that abstains is not yet a release policy, and the rules above differ in how often they release. We therefore compare policies in which an abstention hands the decision to a named fallback, by the strength-and-carbon loss per decision
\begin{equation}
EL=\E[\mathrm{CO_2}]+\lambda_f\,\Prob(\text{strength failure})+\lambda_a\,\Prob(\text{no release})+C_{\mathrm{cal}}\,\ind\{\text{campaign}\}.
\end{equation}
Here CO\textsubscript{2} is the carbon of the released mix in \unitC, taken as zero if nothing is released, and $\lambda_f$ is the penalty for a delivered batch that fails strength. The penalty $\lambda_a=300$ charges the few cells in which a policy has no admissible mix, and $C_{\mathrm{cal}}$ is the cost of the calibration campaign per decision, all in the same units. The released mixes carry about 148 to 274~\unitC\ on average (Table~\ref{tab:loss}), so $\lambda_f=1{,}000$ values a failure at roughly four to seven times the carbon of the mix itself; we scan $\lambda_f$ from 0 to 5,000. Certificate policies use the in-region design of Section~5.2. Two limits apply. The comparison is unrestricted: $\tau$ shapes only the certificate's region, while the other rules and the fallbacks may release any mix, so it does not rank the rules under a common carbon cap. And workability rejections at the truck are not in Eq.~(9), so it is not a full operational evaluation.

\begin{table}[t]
\caption{Expected loss, Eq.~(9), per decision of release policies with a named fallback after abstention, with $\lambda_a=300$ and $C_{\mathrm{cal}}=0$, in \unitC}
\label{tab:loss}
\centering\small
\begin{tabular}{@{}lrrrrr@{}}
\toprule
 & & & \multicolumn{3}{c}{$EL$, $C_{\mathrm{cal}}=0$}\\
\cmidrule(l){4-6}
Policy & $P$(fail) & E[CO\textsubscript{2}] & $\lambda_f=200$ & $\lambda_f=1{,}000$ & $\lambda_f=5{,}000$\\
\midrule
\multicolumn{6}{@{}l}{Laboratory case, $s=0$}\\
Model probability & 0.273 & 164.3 & 219.0 & 437.6 & 1,531.0\\
EN-style margin & 0.000 & 219.6 & 219.6 & 219.6 & 219.6\\
ACI-style margin & 0.000 & 273.6 & 276.6 & 276.6 & 276.6\\
Reuse tested recipe (DEV) & 0.000 & 169.1 & 169.1 & 169.1 & 169.1\\
Reuse tested recipe (campaign)$^{a}$ & 0.000 & 148.5 & 158.9 & 158.9 & 158.9\\
Certificate, else reuse (campaign)$^{a}$ & 0.010 & 162.6 & 175.2 & 183.4 & 224.9\\
Certificate, else EN-style margin$^{a}$ & 0.010 & 222.0 & 224.1 & 232.4 & 273.9\\
Certificate, else ACI-style margin$^{a}$ & 0.010 & 265.2 & 270.2 & 278.5 & 320.0\\
\multicolumn{6}{@{}l}{$s=2$~MPa}\\
Model probability & 0.330 & 164.3 & 230.3 & 494.3 & 1,814.3\\
EN-style margin & 0.010 & 219.6 & 221.6 & 229.6 & 269.6\\
ACI-style margin & 0.000 & 273.6 & 276.6 & 276.6 & 276.6\\
Reuse tested recipe (DEV) & 0.217 & 169.1 & 212.4 & 385.8 & 1,252.4\\
Reuse tested recipe (campaign)$^{a}$ & 0.281 & 149.7 & 216.3 & 440.9 & 1,563.9\\
Certificate, else reuse (campaign)$^{a}$ & 0.252 & 160.4 & 221.3 & 423.0 & 1,431.9\\
Certificate, else EN-style margin$^{a}$ & 0.013 & 223.8 & 226.3 & 236.4 & 286.7\\
Certificate, else ACI-style margin$^{a}$ & 0.003 & 271.3 & 274.9 & 277.3 & 289.1\\
\multicolumn{6}{@{}l}{$s=4$~MPa}\\
Model probability & 0.403 & 164.3 & 245.0 & 567.6 & 2,181.0\\
EN-style margin & 0.090 & 219.6 & 237.6 & 309.6 & 669.6\\
ACI-style margin & 0.013 & 273.6 & 279.3 & 290.0 & 343.3\\
Reuse tested recipe (DEV) & 0.317 & 169.1 & 232.4 & 485.8 & 1,752.4\\
Reuse tested recipe (campaign)$^{a}$ & 0.417 & 147.8 & 241.6 & 575.3 & 2,243.4\\
Certificate, else reuse (campaign)$^{a}$ & 0.391 & 155.5 & 244.3 & 557.5 & 2,123.4\\
Certificate, else EN-style margin$^{a}$ & 0.087 & 223.5 & 240.9 & 310.5 & 658.7\\
Certificate, else ACI-style margin$^{a}$ & 0.016 & 273.5 & 279.6 & 292.1 & 354.3\\
\bottomrule
\end{tabular}

\smallskip
\parbox{\linewidth}{\footnotesize Note: $P$(fail) is the probability that the released mix fails strength on its delivered batch. Certificate policies average over the nine carbon budgets with equal weight; the other rules ignore the budget, so there is no common carbon cap. $^{a}$Policies that run the 40-mix calibration campaign; each of their losses rises by $C_{\mathrm{cal}}$.}
\end{table}

Over $\lambda_f$ from 0 to 5,000, no policy containing the certificate had the smallest expected loss at any level of production variability, even with a free campaign. This holds for every $\lambda_a$ up to 4,739; a certificate policy first ties the best rule at $\lambda_a=4{,}739.06$ (Section~S20 of the supporting information). In the laboratory case, the cheapest recipe that passed during the campaign was best for every $\lambda_f\ge 100$. Its expected loss is 10.2~\unitC\ below that of reusing a development recipe, so the campaign pays for itself if it costs less than that per decision. Under production variability with $\lambda_f\ge 500$, a code-style margin alone was best. Using the certificate first, with the margin as fallback, raised the loss: at $s=4$~MPa its released mixes complied in 0.948 of deliveries, against 0.987 for the ACI-style rule.

In this population the certificate's value is its guaranteed error rate, not a lower expected loss. A population in which the model certifies markedly cheaper mixes than the codes allow could reverse the ranking, and the loss (9) is the calculation that would show it.

\section{Discussion}
\subsection{Recommendations for Practice}
We draw five recommendations for releasing a mix, or any design, on a model prediction.

(i) Validate the model's probability where the release rule acts, not on average. In this study a model that predicted strength with a root mean squared error of 0.9~MPa stated 0.985 at the mix it selected, and that mix complied in 0.727 of replications. Accuracy over the pool does not transfer to the selected mix, because of the optimizer's curse (Section~3). A producer can check this on its own records by comparing the stated probabilities of past releases with their test results.

(ii) When an error rate is needed for the released mix, design the calibration that supplies it. The protocol of Section~4 needs no assumption on the model. In its simplest form it subtracts from the released mix's predicted strength an order statistic of the overpredictions observed on tested mixes in the release region. At $\delta=0.10$ at least nine tested mixes must fall in the region, and the chance of that, and so of a finite bound, can be computed before any mix is cast [Eq.~(8)].

(iii) Calibrate on the population the decision is exposed to. Laboratory calibration certifies laboratory strength. Calibrating on delivered production batches restores the guarantee for what is delivered, at a cost in how often a mix is released, and the certificate's own coverage diagnostics cannot reveal the mismatch (Section~5.2).

(iv) Report the error rate separately from the policy, and compare policies on equal information. A controlled error rate says nothing about how often the rule releases, whether a released mix meets the requirements it does not certify, or what happens after an abstention. Compare policies by a loss such as Eq.~(9), with the fallback named. In this study the policy to beat was the cheapest tested recipe in the laboratory case and a code-style margin under production variability (Section~5.3).

(v) When several dependent requirements are combined into one probability, estimate their dependence and represent it, since the product of marginal probabilities is conservative or optimistic according to its sign (Section~3.2); and when a model also guides the search, read the bottom bin of its reliability diagram, where permanent exclusion shows (Section~S2.6 of the supporting information).

\subsection{Limitations}
The concrete study is retrospective, on 103 laboratory mixes from one laboratory with one strength per mix, and its sources do not state the test specimen. The recorded strengths are close to a smooth function of the mix, so the laboratory case is a low-noise regime (Section~S18 of the supporting information). Production variability is simulated: external records from other concretes bracket its level (Section~S30) but do not measure it on these mixes. The protocol certifies a mix drawn at random within the budget rather than the carbon-minimizing mix, and the code comparisons use margin arithmetic only. The policy comparison uses linear strength-and-carbon losses, without a common carbon cap or the cost of workability rejections. Several comparisons have limited power and are reported as null results, not as evidence of equivalence. The supporting studies of the search, on benchmarks and a remediation simulator, use an isotropic squared-exponential Gaussian process with a grid-chosen length-scale, and the remediation model is an analytic-element idealization.

\subsection{Conclusions}
A model probability of compliance is a ranking device that is also read as a statement of risk. As a ranking device it served well here: the strength model was accurate. As a statement of risk it failed where it was used, at the mix the rule selected: for a fixed population the model is an approximation, and a rule that picks the most favorable mix is exposed to its errors, the optimizer's curse. A designed calibration restores a finite-sample error rate for the released mix, at a cost in how often a mix is released and in testing effort; the effort, and the probability of a finite bound, can be computed in advance. On this population the error rate held at every carbon budget, but no release policy built on it had the smallest expected loss over the penalties examined, and recipes tested during the calibration campaign were often the better choice. The practical lesson is to separate the two roles: let the model choose what to release, attach to the released mix a probability that has been checked against data the choice did not use, and then judge the release policy on its own terms. Potential future directions include repeating the analysis on larger records of practice, such as a contractor's mix-proportion ledger of 719 mixes \citep{lu2024}; running the certificate and the release policies on a plant's production record, where batch-to-batch variability is measured rather than simulated; certifying strength and workability jointly; calibration campaigns designed for the loss (9) rather than for the error rate alone; and dependence models such as copula processes \citep{wilson2010} for several requirements at once.

\section*{Acknowledgments}
{\sloppy The author thanks I-Cheng Yeh, Ke Lu and Ming Qiu, and Farzad Rezazadeh and co-authors for making their data openly available. Claude (Anthropic) assisted with text editing, analysis-code checks and refactoring, replication packaging, and preparation of this version of the manuscript, and Codex (OpenAI) assisted with technical review, checks against archived results, and manuscript shortening. No AI tool generated data or results. The author verified all AI-assisted material and takes full responsibility for the article.\par}

\section*{Funding}
This work was supported by the Anusandhan National Research Foundation (ANRF), India, under the MATRICS scheme (Grant ANRF/ARGM/2025/000532/MTR).

\section*{Conflict of Interest Statement}
The author declares no conflicts of interest.

\section*{Data Availability Statement}
The Concrete Slump Test data \citep{yeh2007a} are openly available from the UCI Machine Learning Repository at \url{https://doi.org/10.24432/C5FG7D}. The repeated-strength records used to bracket production variability are openly available from Mendeley Data \citep{lu2024} at \url{https://doi.org/10.17632/zrsbhndz9f.1} and from the University of Kassel data repository \citep{rezazadeh2026} at \url{https://doi.org/10.48662/daks-56.2}. The code that produces every reported number, table, and figure, the frozen results, the Lean~4 proof files, and copies of all three data sets are provided as supporting information and are also available at \url{https://drive.google.com/drive/folders/1DCjex5HX_PMAr9aw0ybv3pj_RJ1DBHJh}.

\section*{Supporting Information}
Additional supporting information can be found online in the Supporting Information section. The supporting document (Sections~S1--S30, Tables~S1--S26, Figs.~S1--S4) gives the theory of the stated probability of compliance when the model also guides the search, the benchmark and groundwater remediation studies, the prespecified tests, the proofs of all results with their machine-checked Lean~4 counterparts, the external estimates of production variability, and further results and verification studies of the concrete study. Three replication archives contain the code, frozen results, and data needed to reproduce every reported number.

\bibliographystyle{plainnat}
\bibliography{refs}

\begin{thebibliography}{21}
\providecommand{\natexlab}[1]{#1}
\providecommand{\url}[1]{\texttt{#1}}
\expandafter\ifx\csname urlstyle\endcsname\relax
  \providecommand{\doi}[1]{doi: #1}\else
  \providecommand{\doi}{doi: \begingroup \urlstyle{rm}\Url}\fi

\bibitem[{ACI Committee 318}(2011)]{aci318}
{ACI Committee 318}.
\newblock \emph{Building code requirements for structural concrete ({ACI}
  318-11) and commentary}.
\newblock American Concrete Institute, Farmington Hills, MI, 2011.

\bibitem[Anderson and Moncaster(2020)]{anderson2020}
J.~Anderson and A.~Moncaster.
\newblock Embodied carbon of concrete in buildings, {Part} 1: {Analysis} of
  published {EPD}.
\newblock \emph{Buildings and Cities}, 1\penalty0 (1):\penalty0 198--217, 2020.
\newblock \doi{10.5334/bc.59}.

\bibitem[Barber et~al.(2023)Barber, Cand{\`e}s, Ramdas, and
  Tibshirani]{barber2023}
R.~F. Barber, E.~J. Cand{\`e}s, A.~Ramdas, and R.~J. Tibshirani.
\newblock Conformal prediction beyond exchangeability.
\newblock \emph{Annals of Statistics}, 51\penalty0 (2):\penalty0 816--845,
  2023.

\bibitem[{CEN (European Committee for Standardization)}(2021)]{en206}
{CEN (European Committee for Standardization)}.
\newblock Concrete---{Specification}, performance, production and conformity.
\newblock EN 206:2013+A2:2021, CEN, Brussels, Belgium, 2021.

\bibitem[{EFCA (European Federation of Concrete Admixtures
  Associations)}(2015)]{efca2015}
{EFCA (European Federation of Concrete Admixtures Associations)}.
\newblock {EFCA} environmental product declaration: {Concrete}
  admixtures---{Plasticisers} and superplasticisers.
\newblock Model EPD according to ISO 14025 and EN 15804
  EPD-EFC-20150091-IAG1-EN, Institut Bauen und Umwelt, Berlin, 2015.

\bibitem[Fannjiang et~al.(2022)Fannjiang, Bates, Angelopoulos, Listgarten, and
  Jordan]{fannjiang2022}
C.~Fannjiang, S.~Bates, A.~N. Angelopoulos, J.~Listgarten, and M.~I. Jordan.
\newblock Conformal prediction under feedback covariate shift for biomolecular
  design.
\newblock \emph{Proceedings of the National Academy of Sciences}, 119\penalty0
  (43):\penalty0 e2204569119, 2022.
\newblock \doi{10.1073/pnas.2204569119}.

\bibitem[Gardner et~al.(2014)Gardner, Kusner, Xu, Weinberger, and
  Cunningham]{gardner2014}
J.~R. Gardner, M.~J. Kusner, Z.~Xu, K.~Q. Weinberger, and J.~P. Cunningham.
\newblock Bayesian optimization with inequality constraints.
\newblock In \emph{Proc., 31st Int. Conf. on Machine Learning}, pages 937--945.
  PMLR, 2014.

\bibitem[Gelbart et~al.(2014)Gelbart, Snoek, and Adams]{gelbart2014}
M.~A. Gelbart, J.~Snoek, and R.~P. Adams.
\newblock Bayesian optimization with unknown constraints.
\newblock In \emph{Proc., 30th Conf. on Uncertainty in Artificial
  Intelligence}, pages 250--259. AUAI Press, 2014.

\bibitem[Hammond and Jones(2011)]{hammond2011}
G.~P. Hammond and C.~I. Jones.
\newblock \emph{Embodied carbon: {The} inventory of carbon and energy ({ICE})}.
\newblock BSRIA Guide BG 10/2011. BSRIA and University of Bath, Bracknell, UK,
  2011.

\bibitem[Jin and Ren(2025)]{jin2025}
Y.~Jin and Z.~Ren.
\newblock Confidence on the focal: {Conformal} prediction with
  selection-conditional coverage.
\newblock \emph{Journal of the Royal Statistical Society Series B: Statistical
  Methodology}, 87\penalty0 (4):\penalty0 1239--1259, 2025.
\newblock \doi{10.1093/jrsssb/qkaf016}.

\bibitem[Lu and Qiu(2024)]{lu2024}
K.~Lu and M.~Qiu.
\newblock Dataset of compressive strength and slump of normal concrete.
\newblock Dataset Version 1, Mendeley Data, 2024.
\newblock \url{https://doi.org/10.17632/zrsbhndz9f.1}.

\bibitem[{MPA (Mineral Products Association)}(2025)]{mpa2025}
{MPA (Mineral Products Association)}.
\newblock Embodied {CO$_2$e} of {UK} cement, additions and cementitious
  material.
\newblock Fact Sheet 18, Rev. 4, MPA Cement, London, 2025.

\bibitem[Rezazadeh et~al.(2026)Rezazadeh, Abrishambaf, Zimmermann, and
  Kroll]{rezazadeh2026}
F.~Rezazadeh, A.~Abrishambaf, G.~Zimmermann, and A.~Kroll.
\newblock Dataset on the reproducibility of {UHPC} mechanical properties under
  a fixed recipe with controlled production variability.
\newblock \emph{Scientific Data}, 13:\penalty0 1058, 2026.
\newblock \doi{10.1038/s41597-026-07805-z}.

\bibitem[Schonlau et~al.(1998)Schonlau, Welch, and Jones]{schonlau1998}
M.~Schonlau, W.~J. Welch, and D.~R. Jones.
\newblock Global versus local search in constrained optimization of computer
  models.
\newblock In \emph{New developments and applications in experimental design},
  IMS Lecture Notes---Monograph Series 34, pages 11--25. Institute of
  Mathematical Statistics, Hayward, CA, 1998.

\bibitem[Smith and Winkler(2006)]{smith2006}
J.~E. Smith and R.~L. Winkler.
\newblock The optimizer's curse: {Skepticism} and postdecision surprise in
  decision analysis.
\newblock \emph{Management Science}, 52\penalty0 (3):\penalty0 311--322, 2006.
\newblock \doi{10.1287/mnsc.1050.0451}.

\bibitem[Tibshirani et~al.(2019)Tibshirani, Barber, Cand{\`e}s, and
  Ramdas]{tibshirani2019}
R.~J. Tibshirani, R.~F. Barber, E.~J. Cand{\`e}s, and A.~Ramdas.
\newblock Conformal prediction under covariate shift.
\newblock In \emph{Advances in Neural Information Processing Systems 32}, Red
  Hook, NY, 2019. Curran Associates.

\bibitem[Wieczorek(2023)]{wieczorek2023}
J.~Wieczorek.
\newblock Design-based conformal prediction.
\newblock \emph{Survey Methodology}, 49\penalty0 (2), 2023.

\bibitem[Wilson and Ghahramani(2010)]{wilson2010}
A.~G. Wilson and Z.~Ghahramani.
\newblock Copula processes.
\newblock In \emph{Advances in Neural Information Processing Systems 23}, pages
  2460--2468, Red Hook, NY, 2010. Curran Associates.

\bibitem[Yeh(1998)]{yeh1998}
I.-C. Yeh.
\newblock Modeling of strength of high-performance concrete using artificial
  neural networks.
\newblock \emph{Cement and Concrete Research}, 28\penalty0 (12):\penalty0
  1797--1808, 1998.

\bibitem[Yeh(2007{\natexlab{a}})]{yeh2007a}
I.-C. Yeh.
\newblock Concrete slump test.
\newblock Dataset, UCI Machine Learning Repository, 2007{\natexlab{a}}.
\newblock \url{https://doi.org/10.24432/C5FG7D}.

\bibitem[Yeh(2007{\natexlab{b}})]{yeh2007b}
I.-C. Yeh.
\newblock Modeling slump flow of concrete using second-order regressions and
  artificial neural networks.
\newblock \emph{Cement and Concrete Composites}, 29\penalty0 (6):\penalty0
  474--480, 2007{\natexlab{b}}.
\newblock \doi{10.1016/j.cemconcomp.2007.02.001}.

\end{thebibliography}

\end{document}